\documentstyle[aps,pra,multicol,psfig]{revtex}

\begin{document}

\title{Coherent Phenomena in Photonic Crystals}

\author{D. G. Angelakis, E. Paspalakis and P. L. Knight}

\address{Optics Section, Blackett Laboratory,
Imperial College, London SW7 2BZ, United Kingdom}

\date{\today}

\maketitle

\begin{abstract}
We study the spontaneous emission, the absorption and dispersion
properties of a ${\bf \Lambda}$-type atom where one transition
interacts near resonantly with a double-band photonic crystal.
Assuming an isotropic dispersion relation near the band edges, we
show that two distinct coherent phenomena can occur. First, the
spontaneous emission spectrum of the adjacent free space
transition obtains `dark lines' (zeroes in the spectrum). Second,
the atom can become transparent to a probe laser field coupling to
the adjacent free space transition.
\end{abstract}

%\begin{multicols}{1}

\section{Introduction}

It has been well established over the years that spontaneous
emission depends not only on the properties of the excited atomic
system but also on nature of the surrounding environment and more
specifically on the density of electromagnetic vacuum modes.
Purcell \cite{Purcell61a} was the first to predict that the rate
of spontaneous emission for atomic transitions resonant with
cavity frequencies could be enhanced due to the strong frequency
dependence of the density of modes. Employing a similar idea,
Kleppner \cite{Kleppner81a} showed that spontaneous emission in a
waveguide may  be suppressed below the free space level due to the
singular behaviour of the density of modes near the fundamental
threshold frequency of the waveguide.

The above studies attracted recent attention when it was realized
that periodic dielectric structures could be engineered in such
way that gaps in the allowed photon frequencies may appear
\cite{Lambropoulos00a}. Since then, photonic band gap (PBG)
materials exhibiting photon localization have been fabricated
initially at microwave frequencies and more recently at near
infrared and optical frequencies. The study of quantum and
nonlinear optical phenomena in atoms (impurities) embedded in such
structures, lead to the prediction of many interesting effects
\cite{Lambropoulos00a}.
 As examples we
mention the localization of light and the formation of
single-photon \cite{John87a,John90a,John95a,Zhu00c} and two-photon
\cite{Nikolopoulos00a} `photon-atom bound states', suppression and
even complete cancellation of spontaneous emission
\cite{Yablonovitch87a,Quang94a,Kurizki94a,Bay97a}, population
trapping in two-atom systems \cite{Bay97a}, phase dependent
behaviour of the population dynamics
\cite{Quang97a,Woldeyohannes99a}, enhancement of spontaneous
emission interference \cite{Zhu97a,Zhu00a,Zhu00d}, modified
reservoir induced transparency
\cite{Paspalakis99a,Lee99a,Erhard00a} and transient lasing without
inversion \cite{Angelakis00a}, occurrence of dark lines in
spontaneous emission \cite{Paspalakis99b,Zhu00b} and other
phenomena \cite{Bay97b,Vats98a,Nikolopoulos99a}. In addition there
is also current interest with regard to the feasibility of the
observation of either the quantum Zeno effect \cite{Kofman96a} or
the quantum Anti-Zeno effect \cite{Lewenstein99a} in modified
reservoirs, such as the PBG.

In this paper we study the spontaneous emission and the probe
absorption/dispersion spectrum in a ${\bf \Lambda}$-type system,
similar to the one used in previous studies
\cite{Quang94a,Quang97a,Paspalakis99a,Angelakis00a,Paspalakis99b,Bay97b,Lewenstein88a},
with one of the atomic transitions decaying spontaneously in a
double-band \cite{Zhu00d} PBG reservoir. We show that the
spontaneous emission spectrum can exhibit `dark lines' (zeroes in
the spectrum at certain values of the emitted photon frequency) in
the spontaneous emission spectrum in the free space transition. In
addition, the atom becomes transparent to a probe laser field
coupling to the free space transition, which exhibits ultra-slow
group velocities near the transparency window. The dependence of
these phenomena on the width of the gap is examined and a
comparison with the single-band case results is made
\cite{Paspalakis99a,Paspalakisb}.

This article is organized as follows. In section II we apply the
time-dependent Schr\"{o}dinger equation to describe the
interaction of our system with the modified vacuum, calculate the
spontaneous emission spectrum in the free space reservoir and
discuss its properties along with the differences from the
single-band case \cite{Quang94a,Kurizki94a}. In section III using
the time-dependent Schr\"{o}dinger equation we calculate
analytically the steady state linear susceptibility of the system.
Results of the absorption and dispersion of a probe laser field in
this system and their comparison with the single-band case are
also presented in section III. Finally, we summarize our findings
in section IV.

\section{Equations and results for the spontaneous emission spectrum}

We begin with the study of the $\Lambda$-type scheme, shown in
Fig.\ 1(a). This system is similar to that used in previous
studies
\cite{Quang94a,Quang97a,Paspalakis99a,Angelakis00a,Paspalakis99b,Bay97b,Lewenstein88a}.
The atom is assumed to be initially in state $|2\rangle$. The
transition $|2\rangle \leftrightarrow |1\rangle$ is taken to be
near resonant with a modified reservoir (this will be later
referred to as the non-Markovian reservoir), while the transition
$|2\rangle \leftrightarrow |0\rangle$ is assumed to be occurring
in free space (this will be later referred to as the Markovian
reservoir). The spectrum of this latter transition is of central
interest in this section. The Hamiltonian which describes the
dynamics of this system, in the interaction picture and the
rotating wave approximation (RWA), is given by (we use units such
that $\hbar = 1$),
\begin{eqnarray}
H &=& \sum_{{\bf \lambda}}g_{{\bf \lambda}}e^{-i(\omega_{\bf
\lambda}- \omega_{20})t} |2\rangle \langle 0| a_{{\bf \lambda}} +
\sum_{{\bf \kappa}}g_{{\bf \kappa}}e^{-i(\omega_{\bf \kappa}-
\omega_{21})t} |2\rangle \langle 1| a_{{\bf \kappa}} +
\mbox{H.c.}  \, . \label{Ham}
\end{eqnarray}
Here, $g_{{\bf \kappa}}$ denotes the coupling of the atom with the
modified vacuum modes $({\bf \kappa})$ and $g_{{\bf \lambda}}$
denotes the coupling of the atom with the free space vacuum modes
$({\bf \lambda})$. Both coupling strengths are taken to be real.
The energy separations of the states are denoted by $\omega_{ij}
= \omega_{i} - \omega_{j}$ and $\omega_{\bf \kappa}$
$(\omega_{\bf \lambda})$ is the energy of the ${\bf \kappa}$
$({\bf \lambda})$-th reservoir mode.

The description of the system is done via a probability amplitude
approach. We proceed by expanding the wave function of the
system, at a specific time $t$, in terms of the `bare' state
vectors such that
\begin{eqnarray}
 |\psi(t)\rangle = b_{2}(t)|2,\{0\}\rangle
 + \sum_{{\bf \lambda}}b_{{\bf \lambda}}(t)|0,\{{\bf \lambda}\}\rangle
+ \sum_{{\bf \kappa}}b_{{\bf \kappa}}(t)|1,\{{\bf \kappa}\}\rangle
 \, . \label{wav}
\end{eqnarray}
Substituting Eqs.\ (\ref{Ham}) and (\ref{wav}) into the
time-dependent Schr\"{o}dinger equation we obtain
\begin{eqnarray}
i\dot{b}_{2}(t) &=& \sum_{\bf \lambda} g_{{\bf \lambda}} b_{\bf
\lambda} (t) e^{-i(\omega_{{\bf \lambda}} - \omega_{20}) t} +
\sum_{\bf \kappa} g_{{\bf \kappa}} b_{\bf \kappa} (t) e^{-i(\omega_{{\bf \kappa}} - \omega_{21}) t} \label{21} \, , \\
i\dot{b}_{\bf \lambda}(t) &=& g_{{\bf \lambda}} b_{2}(t) e^{i(\omega_{{\bf \lambda}} - \omega_{20}) t} \label{2l} \, , \\
i\dot{b}_{\bf \kappa}(t) &=& g_{{\bf \kappa}} b_{2}(t)
e^{i(\omega_{{\bf \kappa}} - \omega_{21}) t} \label{2k} \, .
\end{eqnarray}
We proceed by performing a formal time integration of Eqs.\
(\ref{2l}) and (\ref{2k}) and substitute the result into Eq.\
(\ref{21}) to obtain the integro-differential equation
\begin{eqnarray}
\dot{b}_{2}(t) &=& -\int^{t}_{0}dt^{\prime}b_{2}
(t^{\prime})\sum_{\bf \lambda} g^{2}_{{\bf \lambda}}
e^{-i(\omega_{{\bf \lambda}} - \omega_{20}) (t-t^{\prime})} -
\int^{t}_{0}dt^{\prime}b_{2} (t^{\prime})\sum_{\bf \kappa}
g^{2}_{{\bf \kappa}}  e^{-i(\omega_{{\bf \kappa}} - \omega_{21})
(t-t^{\prime})}  \label{21a} \, .
\end{eqnarray}
Because the reservoir with modes $\lambda$ is assumed to be
Markovian, we can apply the usual Weisskopf-Wigner result
\cite{Lambropoulos00a} and obtain
\begin{equation}
\sum_{\bf \lambda} g^{2}_{{\bf \lambda}}  e^{-i(\omega_{{\bf
\lambda}} - \omega_{20}) (t-t^{\prime})} =
\frac{\gamma}{2}\delta(t - t^{\prime}) \, . \label{weisskopf}
\end{equation}
Note that the principal value term associated with the Lamb shift
which should accompany the decay rate term has been omitted in
Eq.\ (\ref{weisskopf}). This does not affect our results, as we
can assume that the Lamb shift is incorporated into the
definition of our state energies. For the second summation in
Eq.\ (\ref{21a}), the one associated with the modified reservoir
modes, the above result is not applicable as the density of modes
of this reservoir is assumed to vary much quicker than that of
free space. To tackle this problem, we define the following kernel
\begin{equation}
K(t-t^{\prime}) = \sum_{{\bf \kappa}}  g^{2}_{{\bf \kappa}}
e^{-i(\omega_{\bf \kappa}- \omega_{21}) (t-t^{\prime})} \approx
\beta^{3/2}\int d\omega \rho(\omega) e^{-i(\omega - \omega_{21})
(t-t^{\prime})} \, , \label{kerneltot}
\end{equation}
with $\beta$ being the atom-modified reservoir resonant coupling
constant. The above kernel is calculated using the appropriate
density of modes $\rho(\omega)$ of the modified reservoir. Using
Eqs.\ (\ref{weisskopf}) and (\ref{kerneltot}) into Eq.\
(\ref{21a}) we obtain
\begin{eqnarray}
\dot{b}_{2}(t) &=& -\frac{\gamma}{2} b_{2} (t) -
\int^{t}_{0}dt^{\prime}b_{2} (t^{\prime}) K(t-t^{\prime})
\label{21a1} \, .
\end{eqnarray}

The long time spontaneous emission spectrum in the Markovian
reservoir is given by $S(\delta_{\bf \lambda}) \propto |b_{\bf
\lambda}(t \rightarrow \infty)|^2$, with $\delta_{\bf \lambda} =
\omega_{\bf \lambda} - \omega_{20}$. We calculate $b_{\bf
\lambda}(t \rightarrow \infty)$ with the use of the Laplace
transform of the equations of motion. Using Eq.\ (\ref{2l}) and
the final value theorem we obtain the spontaneous emission
spectrum as $S(\delta_{{\bf \lambda}}) \propto {\gamma} |\lim_{s
\rightarrow -i \delta_{{\bf \lambda}}} B_{2}(s)|^{2}$, where
$B_{2}(s) = \int^{\infty}_{0}dt e^{-st}b_{2}(t)$ is the Laplace
transform of the atomic amplitudes $b_{2}(t)$ and $s$ is the
Laplace variable. This in turn, with the help of Eq.\
(\ref{21a1}), reduces to
\begin{equation}
S(\delta_{{\bf \lambda}}) \propto \frac{\gamma}{\left|-i
\delta_{\bf \lambda} + \gamma/2 + \tilde{K}(s \rightarrow -i
\delta_{\bf \lambda})\right|^2} \, . \label{sp}
\end{equation}
Here, $\tilde{K}(s) = \int^{\infty}_{0}dt e^{-st}K(t)$ is the
Laplace transform of $K(t)$, which yields from Eq.\
(\ref{kerneltot}),
\begin{equation}
\tilde{K}(s) =  \beta^{3/2} \int d\omega \frac{\rho(\omega)}{s + i
(\omega - \omega_{21})} \, . \label{kernellap}
\end{equation}
Therefore, in order to calculate the spontaneous emission
spectrum in the Markovian reservoir we need to calculate
$\tilde{K}(s)$.
%This will be done for different models of the
%density of modes $\rho(\omega)$ of the non-Markovian reservoir.

%In the first model we consider the usual single-band non-Markovian
%reservoir which is obtained near the edge of an isotropic PBG
%model \cite{Quang94a,Kurizki94a}. The near the edge dispersion
%relation is approximated by $\omega_{\kappa} = \omega_{g} + A
%(\kappa - \kappa_{0})^2$, with $A \approx
%\omega_{g}/\kappa_{0}^2$. Then the density of modes reads,
%\begin{equation}
%\rho(\omega) = \frac{1}{\pi} \frac{1}{\sqrt{\omega -
%\omega_{g}}}\Theta(\omega - \omega_{g}) \, , \label{isodom}
%\end{equation}
%where $\omega_{g}$ is the gap frequency and $\Theta$ is the
%Heaviside step function (as shown in Fig.\ 1(b)). In this case,
%Eq.\ (\ref{kernellap}) leads to
%\begin{equation}
%\tilde{K}(s) = \frac{\beta^{3/2}}{\sqrt{i}\sqrt{s + i\delta_{g}}}
%\, , \label{isoker}
%\end{equation}
%with $\delta_{g} = \omega_{g} - \omega_{21}$.

We consider the case of a double-band isotropic model of the
photonic crystal, which has an upper band, a lower band and a
forbidden gap, the near the edges dispersion relation
\cite{Zhu00d} is approximated by
\begin{eqnarray}
\omega_{\kappa}=\omega_{g1}-A_{1}(\kappa-\kappa_{0})^2 \  &,&
\, \kappa < \kappa_{0} \nonumber \\
\omega_{\kappa}=\omega_{g2}+A_{2}(\kappa-\kappa_{0})^2 \  &,& \,
\kappa > \kappa_{0} \, ,
\end{eqnarray}
with $A_{n} \approx \omega_{gn}/ {\kappa}_{0}^2$, $(n=1,2)$. Then
the density of modes reads,
\begin{equation} \rho(\omega) = \frac{1}{2\pi} \left[\frac{1}{\sqrt{\omega_{g1} - \omega}}\Theta(\omega_{g1} -
\omega)+ \frac{1}{\sqrt{\omega - \omega_{g2}}}\Theta(\omega -
\omega_{g2}) \right] \, , \label{2isodom}
\end{equation}
where $\omega_{gn}$ is the gap frequency and $\Theta$ is the
Heaviside step function (as shown in Fig.\ 1(c)). Then, from Eq.\
(\ref{kernellap}) we obtain
\begin{equation}
\tilde{K}(s) = \frac{1}{2}\left(
\frac{\beta^{3/2}\sqrt{i}}{\sqrt{s + i\delta_{g1}}} +
\frac{\beta^{3/2}}{\sqrt{i}\sqrt{s + i\delta_{g2}}} \right) \, ,
\label{2isoker}
\end{equation}
with $\delta_{gn} = \omega_{gn} - \omega_{21}$.

We use the formulae obtained above and calculate the spontaneous
emission for several parameters of the system.
%First we study the
%single-band PBG case. From Eqs.\ (\ref{sp}) and (\ref{isoker}) is
%obvious that the spectrum exhibits a zero (i.e. predicts the
%existence of a dark line), if $\delta_{\bf \lambda} = \delta_{g}$.
%This is purely an effect of the above density of modes of Eq.\
%(\ref{isoker}), and the non-Markovian character of the reservoir.
%In the case of a Markovian reservoir the spontaneous emission
%spectrum would obtain the well-known Lorentzian profile and no
%dark line would appear in the spectrum. The behaviour of the
%spectrum is shown in Fig.\ 2 for different values of the detuning
%from the threshold. The spectrum has two well-separated peaks and
%the dark line appears at the predicted value.
From Eqs.\ (\ref{sp}) and (\ref{2isoker}) is obvious that the
spectrum exhibits two zeros (i.e. predicts the existence of two
dark lines), at $\delta_{\lambda} = \delta_{g1}$ and
$\delta_{\lambda} = \delta_{g2}$. This is purely an effect of the
density of modes of Eq.\ (\ref{2isoker}), and the non-Markovian
character of the reservoir. In the case of a Markovian reservoir
the spontaneous emission spectrum would obtain the well-known
Lorentzian profile and no dark line would appear in the spectrum.
The behaviour of the spectrum is shown in Fig.\ 2. In Fig.\ 2(a)
we show an asymmetric case where the right hand side (rhs)
threshold detuning $(\delta_{g2})$ is kept constant while the left
hand side (lhs) threshold detuning $(\delta_{g1})$ is changed.
There are three peaks and two zeroes in the spectrum, one
pronounced peak in the center and two lower peaks in the sides. As
$\delta_{g1}$ goes further to the sides the lhs peak is suppressed
and the spectrum approximates that of the single band
case\cite{Quang94a,Kurizki94a} which is shown in Fig. 3. That is
because the lhs singularity in the density of the reservoir modes,
Fig 1.\, is far detuned from the atomic transition.

 %We then use Eqs.\ (\ref{sp}) and (\ref{2isoker}) and
%calculate the behaviour of the system in the double-band PBG case.
%It is clear that the spectrum obtains two zeroes at
%$\delta_{\lambda} = \delta_{g1}$ and $\delta_{\lambda} =
%\delta_{g2}$, i.e. two dark lines exist in the spectrum. We
%present the spectrum for the double-band case in Fig.\ 3. In Fig.\
%3(a) we show an asymmetric case where the right hand side (rhs)
%threshold detuning $(\delta_{g2})$ is kept constant while the left
%hand side (lhs) threshold detuning $(\delta_{g1})$ is changed.
%There are three peaks and two zeroes in the spectrum, one
%pronounced peak in the center and two lower peaks in the sides. As
%$\delta_{g1}$ goes further to the sides the lhs peak is suppressed
%and the spectrum approximates that of Fig. 2 (full curve).
In Fig.\ 2(b) we show a symmetric spectrum, which is the case when
the atom is placed in the middle of the gap. This spectrum also
has two zeroes and three peaks. In this case when the threshold
detunings increase (i.e, the width of the gap increases) the
spectrum approximates the Lorentzian shape. This is a result of
the complete cancelation of any emission in the modified reservoir
as there are no modes available for a large range of frequencies
around the relevant atomic transition.
%%We note that in the case of
%%when both transitions occured in free space, the spectrum under
%%consideration would still have a Lorentzian shape but the area
%%underneath the curve would have to be smaller as photons are
%%emitted from the second transition as well \cite{Paspalakis99b}.

\section{Equations and results for the probe absorption/dispersion spectrum}

The aim in this section is to investigate the absorption and
dispersion properties of our system for a {\it weak} probe laser
field. As the probe laser field is assumed to be weak the
probability amplitude approach can be employed for the
description of the system. The Hamiltonian of the system, in the
interaction picture and the rotating wave approximation, is given
by,
\begin{eqnarray}
H &=& \Omega e^{i \delta t} |0\rangle \langle 2| + \sum_{{\bf
\lambda}}g_{{\bf \lambda}}e^{-i(\omega_{\bf \lambda}-
\omega_{20})t} |2\rangle \langle 0| a_{{\bf \lambda}} +
\sum_{{\bf \kappa}}g_{{\bf \kappa}}e^{-i(\omega_{\bf \kappa}-
\omega_{21})t} |2\rangle \langle 1| a_{{\bf \kappa}} +
\mbox{H.c.}  \, . \label{Hama}
\end{eqnarray}
Here, $\Omega$ is the Rabi frequency (assumed real for simplicity)
and $\delta = \omega - \omega_{20}$ is the laser detuning from
resonance with the $|0\rangle \leftrightarrow |2\rangle$
transition with $\omega$ being the angular frequency of the probe
laser field. We note that we are interested in the perturbative
behaviour of the system to the probe laser pulse, therefore we can
eliminate the Markovian decay modes using the method presented in
the previous section and study the system using the following
effective Hamiltonian
\begin{eqnarray}
H = \bigg[ \Omega e^{i \delta t} |0\rangle \langle 2| +
\sum_{\kappa}g_{\kappa}e^{-i(\omega_{\kappa}- \omega_{21})t}
|2\rangle \langle1| a_{\kappa}  + \mbox{H.c.} \bigg] -i
\frac{\gamma}{2} |2\rangle \langle2| \, . \label{Ham1a}
\end{eqnarray}

The wavefunction of the system, at a specific time $t$, can be
expanded in terms of the `bare' eigenvectors such that
\begin{eqnarray}
 |\psi(t)\rangle = c_{0}(t)|0,\{0\}\rangle + c_{2}(t) e^{-i \delta t} |2,\{0\} \rangle +
  \sum_{{\bf \kappa}}c_{{\bf \kappa}}(t)|1,\{{\bf \kappa}\}\rangle \, , \label{wav1}
\end{eqnarray}
and $c_{0}(t=0)=1$, $c_{2}(t=0)=0$, $c_{\kappa}(t=0)=0$.
Substituting Eqs.\ (\ref{Ham1a}) and (\ref{wav1}) into the
time-dependent Schr\"{o}dinger equation and eliminating the vacuum
amplitude $c_{\kappa}(t)$, we obtain the time evolution of the
probability amplitudes as
\begin{eqnarray}
i\dot{c}_{0}(t) &=& \Omega c_{2}(t) \label{b0a} \, ,\\
i\dot{c}_{2}(t) &=& \Omega c_{0}(t) - \left(\delta + i
\frac{\gamma}{2}\right)c_{2}(t)  -i
\int^{t}_{0}dt^{\prime}K^{\prime}(t-t^{\prime})c_{2}(t^{\prime})
\, , \label{b1a}
\end{eqnarray}
with the kernel
\begin{eqnarray}
K^{\prime}(t-t^{\prime}) &=& \sum_{\kappa}  g^{2}_{\kappa}
e^{-i(\omega_{\kappa}- \omega_{21}-\delta) (t-t^{\prime})} \approx
K(t)e^{i \delta t} \, .
 \label{kerneltot1}
\end{eqnarray}
 All the coupling  constants ($g_{\kappa}$, $\beta$,
$\Omega$) are assumed to be real, for simplicity.

The equation of motion for the electric field amplitude $E(z,t)$
is given by \cite{Harris92a},
\begin{eqnarray}
\left( \frac{\partial }{\partial z} +
\frac{1}{v_{g}}\frac{\partial }{\partial t} \right) E(z,t) = - i
\frac{\omega}{2c} \chi(\delta) E(z,t) \, ,
\end{eqnarray}
where $\chi(\delta)$ is the linear susceptibility of the medium
and $v_{g} = c/[1 + (\omega/2)(\partial \mbox{Re}(\chi)/\partial
\omega)]$ is the group velocity of the laser pulse with the
derivative of the real part of the susceptibility being evaluated
at the carrier frequency.

Since the transition $|0\rangle \leftrightarrow |2\rangle$ is
treated as occurring in free space, the steady state linear
susceptibility is given by
\begin{equation}
\chi(\delta) = - \frac{4\pi {\cal N}
|\mbox{\boldmath${\mu}$\unboldmath $_{02}$}|^{2}}{\Omega(z,t)}
c_{0}(t \rightarrow \infty) c^{*}_{2}(t \rightarrow \infty) \,
,\label{suscep}
\end{equation}
with ${\cal N}$ being the atomic density.

Therefore, in order to determine the steady state absorption
properties we have to solve Eqs.\ (\ref{b0a}), (\ref{b1a}) for
long times. We assume that the laser-atom interaction is very weak
$(\Omega \ll \beta, \gamma)$ so that $c_{0}(t) \approx 1$ for all
times, and by using perturbation theory, Eqs. (\ref{b0a}) and
(\ref{b1a}) reduce to
\begin{eqnarray}
i\dot{c}_{2}(t) &\approx& \Omega  - \left(\delta + i
\frac{\gamma}{2}\right)c_{2}(t) - i
\int^{t}_{0}dt^{\prime}K^{\prime}(t-t^{\prime})c_{2}(t^{\prime})
\, . \label{b1approx}
\end{eqnarray}
We further assume that $\Omega(z,t)$ is approximately constant in
the medium and with the use of the Laplace transform we obtain
from Eq.\ (\ref{b1approx})
\begin{equation}
{C}_{2}(s) = \frac{\Omega}{s\left[\delta +i\gamma/2 +
i\tilde{K}(s) + is \right]} \, . \label{A1}
\end{equation}
If $\gamma \neq 0$ then the terms inside the brackets of Eq.
(\ref{A1}) have only complex, not purely imaginary, roots.
Therefore we can easily obtain, using the final value theorem, the
long time behaviour of the probability amplitude,
\begin{eqnarray}
c_{2}(t \rightarrow \infty) &=& \lim_{s \rightarrow 0}[C_{2}(s)]
\nonumber \\ &=& \frac{\Omega}{\delta + i \gamma/2 + i
\tilde{K^{\prime}}(s \rightarrow 0)} \nonumber \\ &=&
\frac{\Omega}{\delta + i \gamma/2 + i \tilde{K}(-i \delta)} \, .
\label{b2fin}
\end{eqnarray}
Using Eqs.  (\ref{suscep}) and (\ref{b2fin}) we find that the
steady state linear susceptibility of our system is given by
\begin{equation}
\chi(\delta) = - \frac{4\pi {\cal N}
|\mbox{\boldmath${\mu}$\unboldmath $_{02}$}|^{2}}{\delta - i
\gamma/2 - i \tilde{K^{*}}(-i \delta)} \, .\label{suscepfin}
\end{equation}
Therefore, the absorption and dispersion of the probe laser field
are determined by the density of modes of the non-Markovian
reservoir. In our case the linear susceptibility becomes zero at
$\delta = \delta_{g1}$ and $\delta = \delta_{g2}$, therefore the
medium becomes transparent to the laser field.  This result is in
contrast with the case in which the transition $|2\rangle
\leftrightarrow |1\rangle$ occurs in free space, where the
well-known Lorentzian absorption profile is obtained. Typical
spectra are shown in Fig.\ 4, where only the lhs threshold
detuning is changed. Both the absorption and dispersion spectra
are asymmetric and their shape depends critically on the
detunings. We note that Fig.\ 4(c) resembles as expected, the
corresponding spectra in the case of a single band reservoir
\cite{Paspalakis99a} which is shown in Fig.\ 5.

We also present the case of symmetric spectra in Fig.\ 6, where
the detunings are symmetrically changed i.e., with the transition
frequency of the atom in the middle of the gap. Here the
interesting effect is best shown in Figs.\ 6(a) and 6(b) where the
transparency effect in the sides of the spectra is combined with
the usual absorption/dispersion profile near the center of the
spectrum.

We note that besides the transparency effect which occurs for the
predicted values of frequency, we note that the group velocity of
the probe laser pulse is also effectively reduced near the
transparency window due to the steepness of the dispersion curve.
These results are related to that of electromagnetically induced
transparency (EIT) in a four-level laser-driven tripod scheme
\cite{Paspalakis00a}. However, EIT occurs through the application
of external laser fields. Here, transparency is intrinsic to the
system as it occurs due to the presence of a square root
singularity of the density of modes at the band edges.

\section{Summary}
In this article we studied the spontaneous emission, absorption
and dispersion spectra of a three-level atom embedded in a
two-band isotropic photonic crystal. For the spontaneous emission
we have shown that the spectrum exhibits two dark lines and can be
significantly modified via the system parameters. The explicit
dependance of the spectrum on the width of the gap and on various
values of the atomic parameters was also analyzed. For the
absorption and dispersion spectra, two transparency windows exist.
Near the transparency windows a very slow group velocity for the
laser pulse is obtained. In this case too, the spectra depend
critically on the atomic parameters and the width of the gap.

\section*{Acknowledgments}
We would like to acknowledge the financial support of the UK
Engineering and Physical Sciences Research Council (EPSRC), the
Hellenic State Scholarship Foundation (SSF) and the European
Commission TMR Network on Microlasers and Cavity QED.

%\end{multicols}
%\pagebreak

\pagebreak

\begin{figure}
\centerline{\psfig{figure=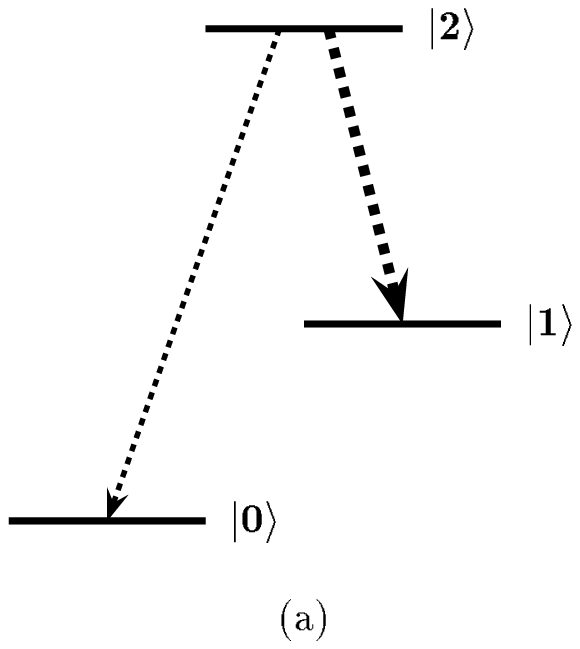,width=4.cm}}

\vspace*{1.mm}

\centerline{\psfig{figure=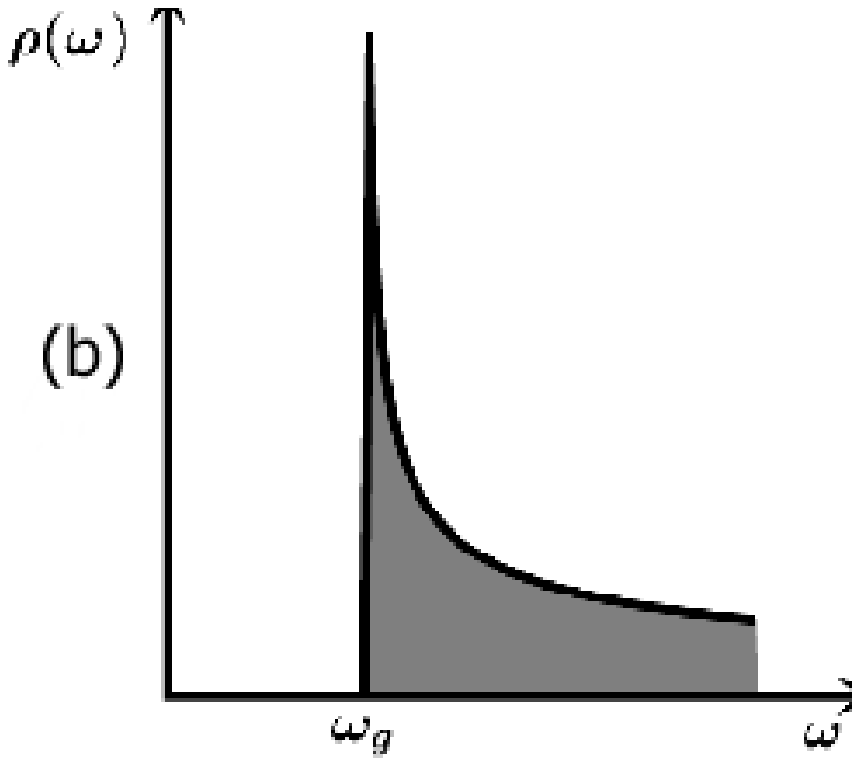,height=7.cm}}
\vspace*{1.mm}

\centerline{\psfig{figure=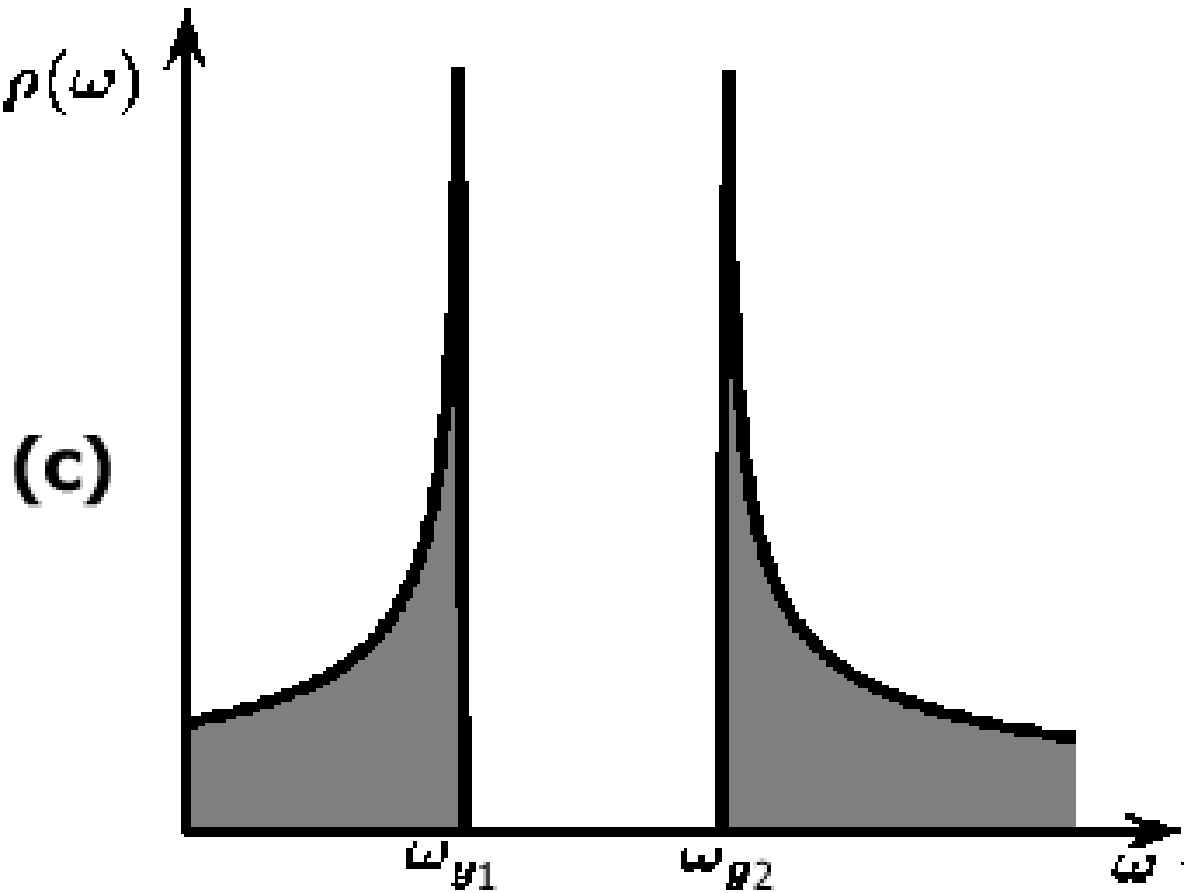,width=8.cm}}

\centerline{ \caption {\narrowtext Figure (a) displays a three
level, ${\bf \Lambda}$-type atomic system. The thick dashed line
denotes the coupling to the modified reservoir (PBG) and the thin
dashed line denotes the background decay. Figure (b) shows the
density of modes for the case of the single-band isotropic PBG
model. Figure (c) shows the density of modes for the case of the
double-band isotropic PBG model.}} \label{Fig1}
\end{figure}

\pagebreak

\begin{figure}
\centerline{\psfig{figure=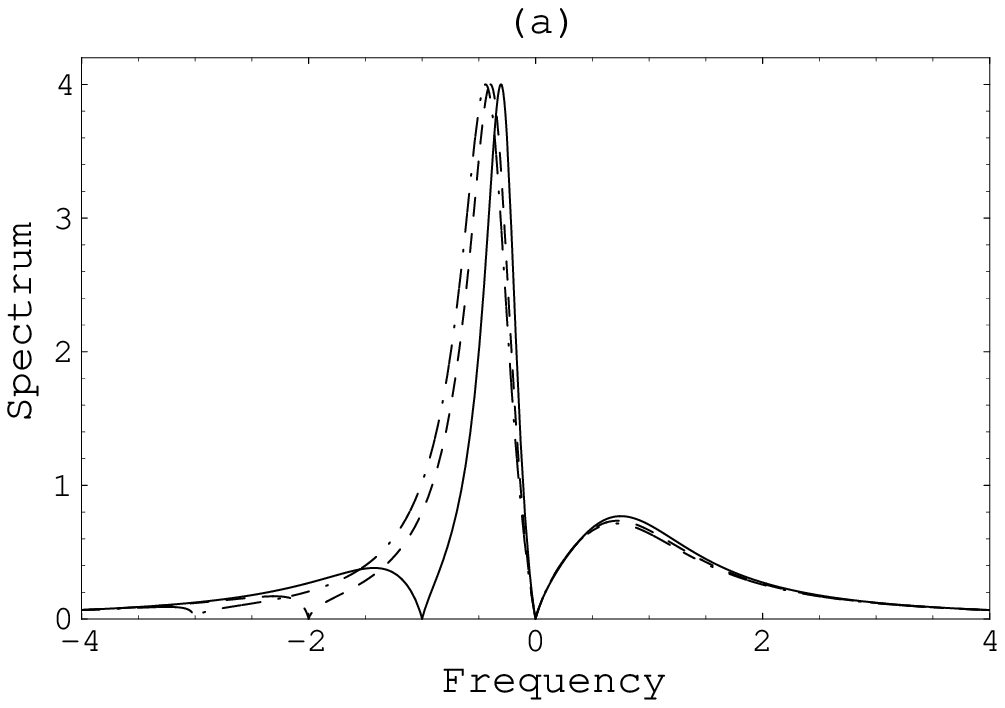,height=5.5cm}}

\vspace*{0.5cm}

\centerline{\psfig{figure=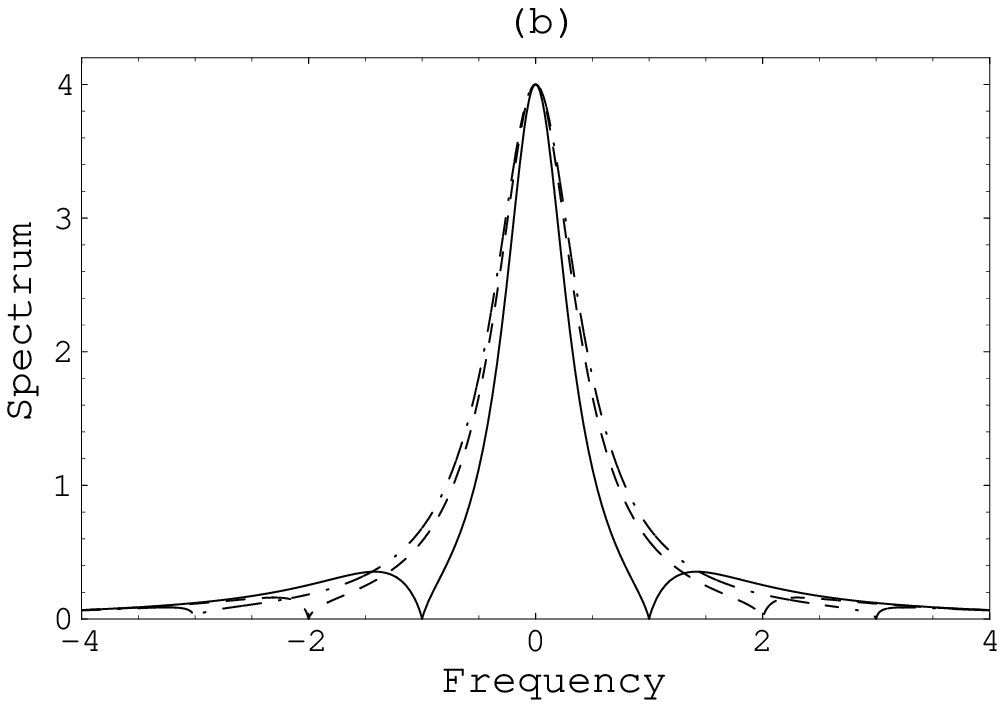,height=5.5cm}} \centerline{
\caption {\narrowtext The spontaneous emission spectrum
$S(\delta_{\bf \lambda})$ (in arbitrary units) for the double-band
case and parameters $\gamma = 1$ and (a) ($\delta_{g1} = -1$,
$\delta_{g2} = 0$) (full curve), ($\delta_{g1} = -2$, $\delta_{g2}
= 0$) (dashed curve), ($\delta_{g1} = -3$, $\delta_{g2} = 0$)
(dot-dashed curve), (b) ($\delta_{g1} = -1$, $\delta_{g2} = 1$)
(full curve), ($\delta_{g1} = -2$, $\delta_{g2} = 2$) (dashed
curve), ($\delta_{g1} = -3$, $\delta_{g2} = 3$) (dot-dashed
curve).}} \label{Fig3}
\end{figure}

\pagebreak

\begin{figure}
\centerline{\psfig{figure=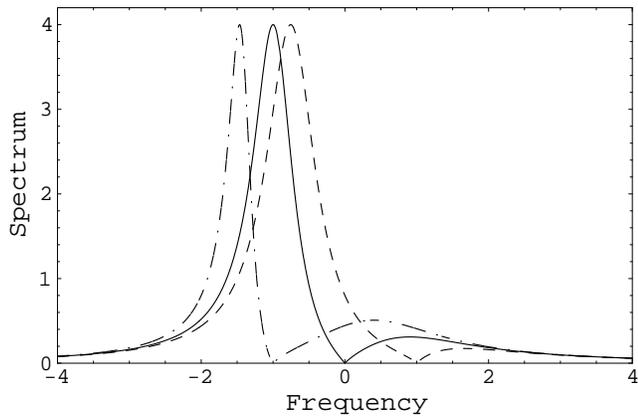,height=5.5cm}} \centerline{
 \caption {\narrowtext The spontaneous emission spectrum
$S(\delta_{\bf \lambda})$ (in arbitrary units) for the single-band
case and parameters $\gamma = 1$ and $\delta_{g} = 0$ (full
curve); $\delta_{g} = 1$ (dashed curve); $\delta_{g} = -1$
(dot-dashed curve). All parameters are in units of $\beta$. }}
\label{Fig2}
\end{figure}

\pagebreak

\begin{figure}
\centerline{\psfig{figure=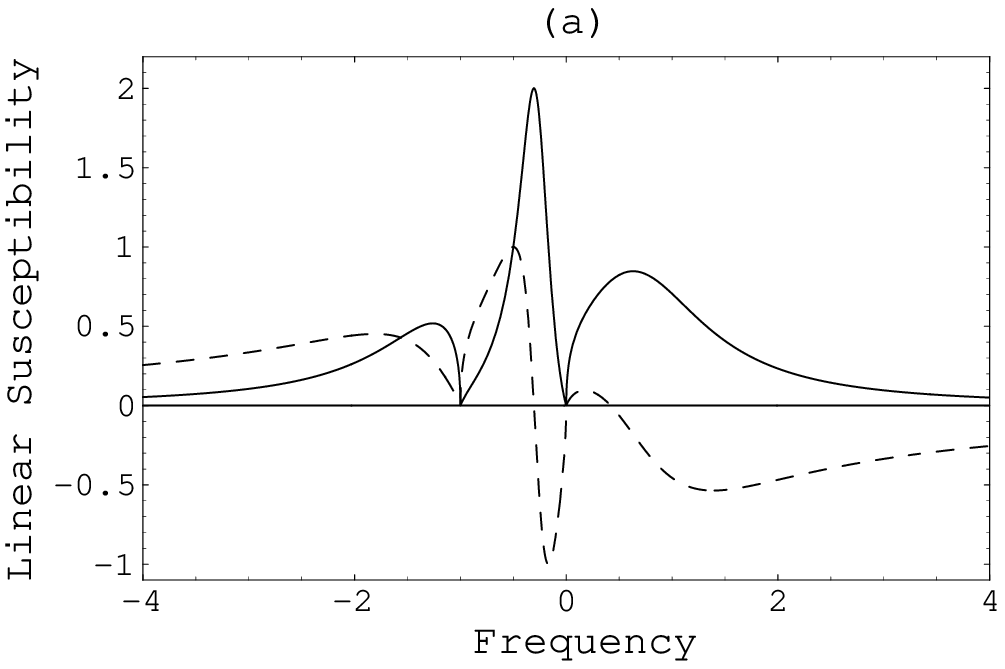,height=5.5cm}}

\vspace*{0.5cm}

\centerline{\psfig{figure=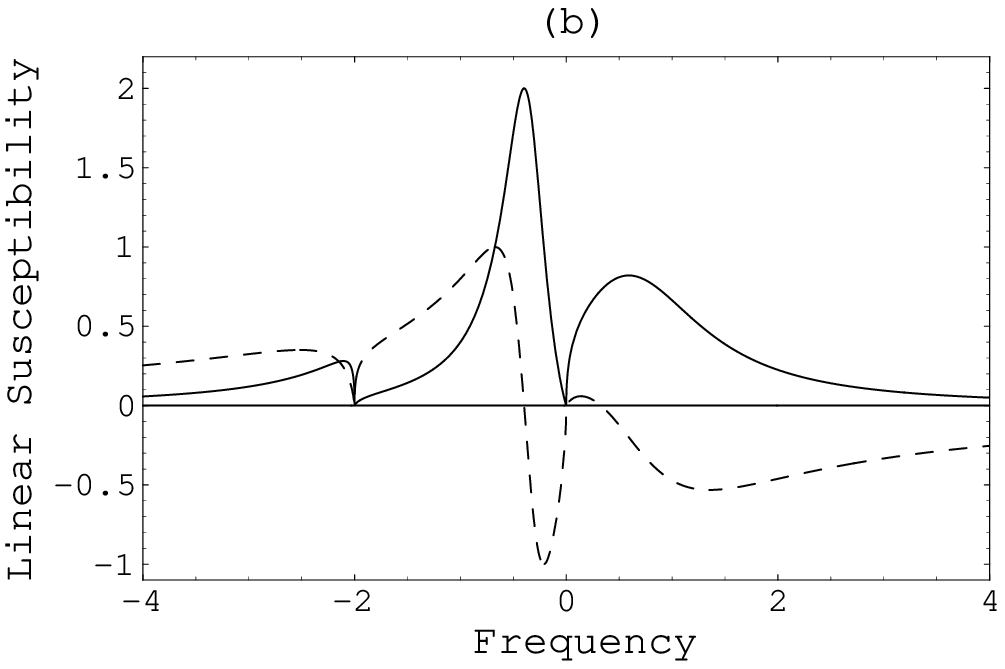,height=5.5cm}}

\vspace*{0.5cm}

\centerline{\psfig{figure=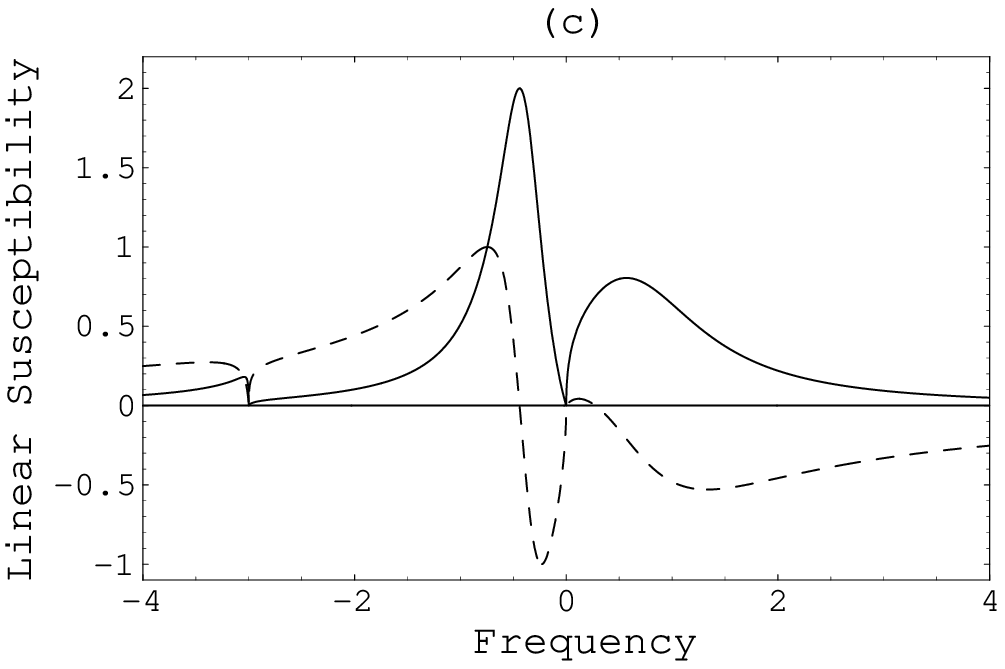,height=5.5cm}}

\centerline{ \caption {\narrowtext  The absorption and dispersion
spectra (in arbitrary units) for the double-band case for
parameters (a) ($\delta_{g1} = -1$, $\delta_{g2} = 0$), (b)
($\delta_{g1} = -2$, $\delta_{g2} = 0$), (c) ($\delta_{g1} = -3$,
$\delta_{g2} = 0$).}} \label{Fig5}
\end{figure}

\pagebreak

\begin{figure}
\centerline{\psfig{figure=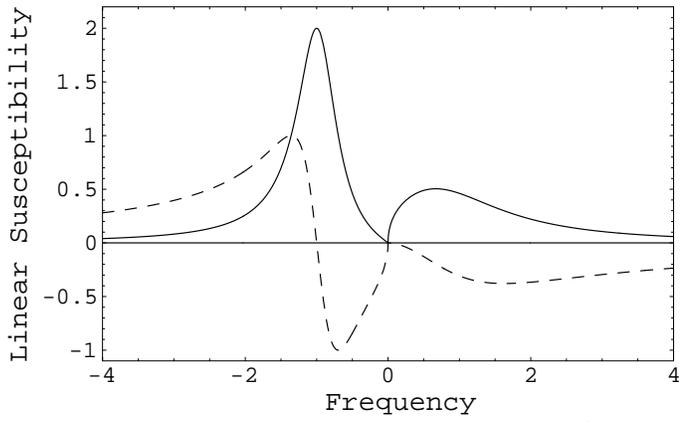,height=5.5cm}} \centerline{
\caption {\narrowtext The absorption and dispersion spectra (in
arbitrary units) for the single-band case for parameters $\gamma =
1$, $\delta_{g} = 0$. The solid curve is the absorption profile
($-$Im$[\chi(\delta)]$) while the dashed curve the dispersion
profile (Re$[\chi(\delta)]$).}} \label{Fig4}
\end{figure}

\pagebreak

\begin{figure}
\centerline{\psfig{figure=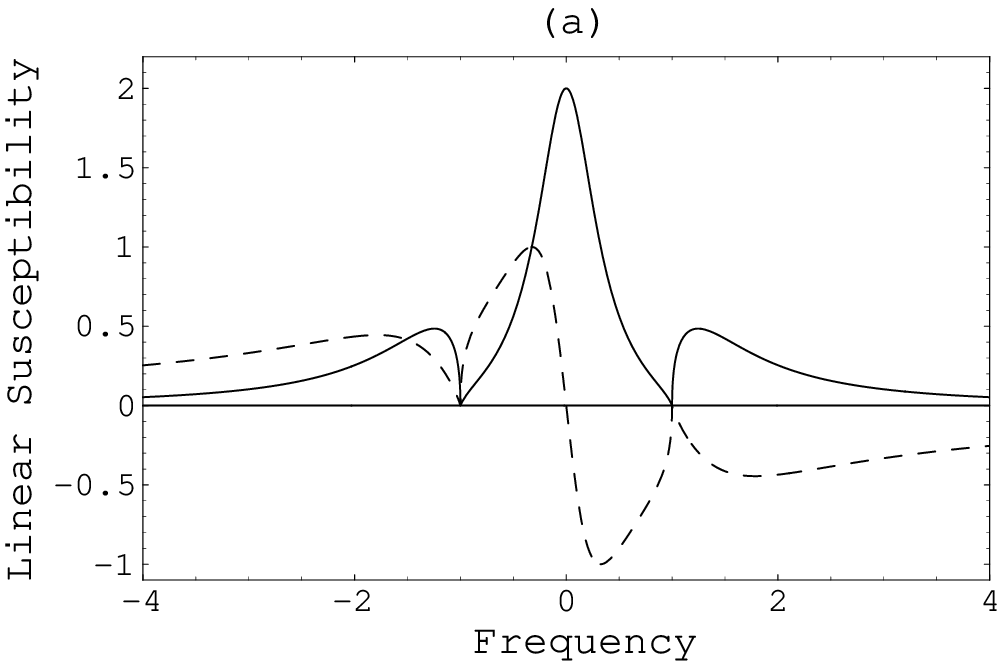,height=5.5cm}}

\vspace*{0.5cm}

\centerline{\psfig{figure=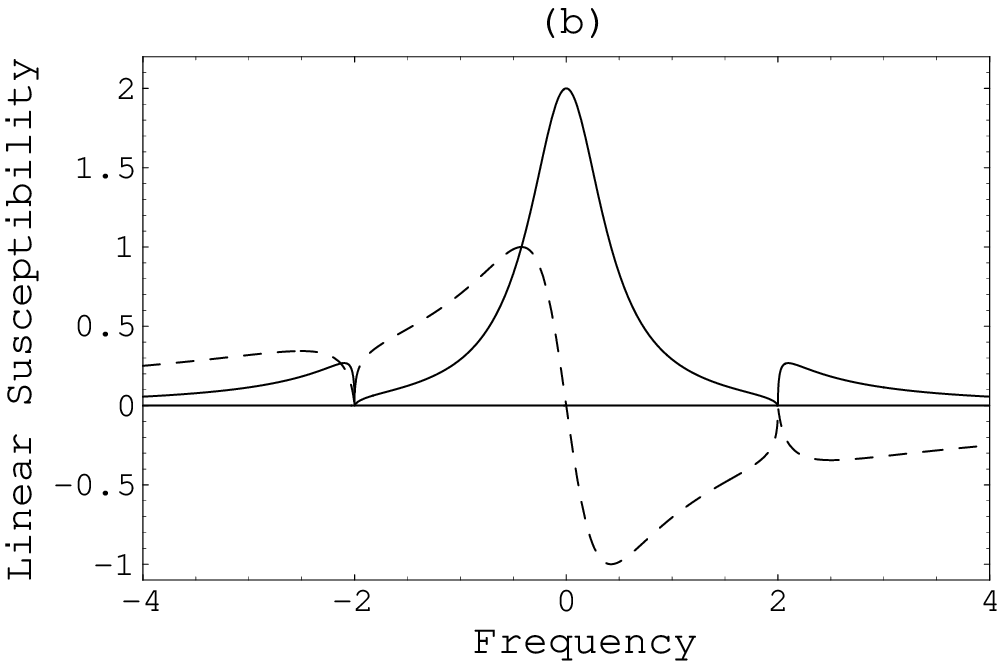,height=5.5cm}}

\vspace*{0.5cm}

\centerline{\psfig{figure=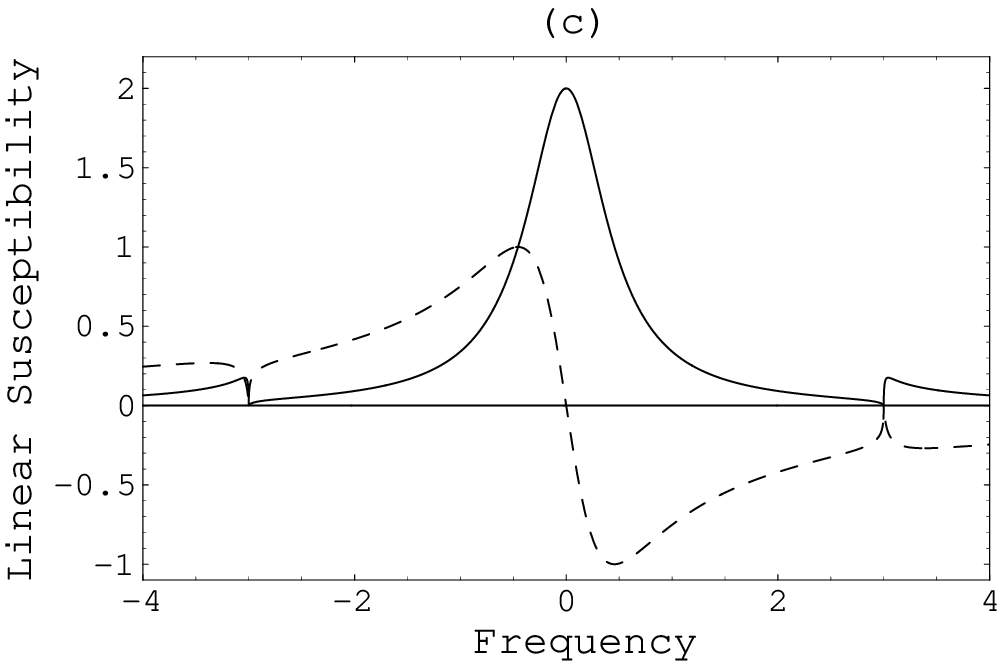,height=5.5cm}}

\centerline{ \caption {\narrowtext The absorption and dispersion
spectra (in arbitrary units) for the double-band case for
parameters (a) ($\delta_{g1} = -1$, $\delta_{g2} = 1$), (b)
($\delta_{g1} = -2$, $\delta_{g2} = 2$), (c) ($\delta_{g1} = -3$,
$\delta_{g2} = 3$).}} \label{Fig6}
\end{figure}

\begin{thebibliography}{99}

\bibitem{Purcell61a}
E.M. Purcell, Phys. Rev. {\bf 69}, 681 (1946).

\bibitem{Kleppner81a}
D. Kleppner, Phys. Rev. Lett. {\bf 47}, 233 (1981).

\bibitem{Lambropoulos00a}
For a recent review in the subject see, P. Lambropoulos, G.M.
Nikolopoulos, T.R. Nielsen and S. Bay, Rep. Prog. Phys. {\bf 63},
455 (2000).

\bibitem{John87a}
S. John, Phys. Rev. Lett. {\bf 58}, 2486 (1987).

\bibitem{John90a}
S. John and J. Wang, Phys. Rev. Lett. {\bf 64}, 2418 (1990);
Phys. Rev. B {\bf 43}, 12772 (1990).

\bibitem{John95a}
S. John and T. Quang, Phys. Rev. Lett. {\bf 74}, 3419 (1995).

\bibitem{Zhu00c}
S.-Y. Zhu, Y.P. Yang, H. Chen, H. Zheng and M.S. Zubairy, Phys.
Rev. Lett. {\bf 84}, 2136 (2000).

\bibitem{Nikolopoulos00a}
G.M. Nikolopoulos and P. Lambropoulos, Phys. Rev. A {\bf 61},
053812 (2000)

\bibitem{Yablonovitch87a}
E. Yablonovitch, Phys. Rev. Lett. {\bf 58}, 2059 (1987).

\bibitem{Quang94a}
S. John and T. Quang, Phys. Rev. A {\bf 50}, 1764 (1994).

\bibitem{Kurizki94a}
A.G. Kofman, G. Kurizki and B. Sherman, J. Mod. Opt. {\bf 41},
353 (1994).

\bibitem{Bay97a}
S. Bay, P. Lambropoulos and K. M{\o}lmer, Phys. Rev. A {\bf 55},
1485 (1997).

\bibitem{Quang97a}
T. Quang, M. Woldeyohannes, S. John and G.S. Agarwal, Phys. Rev.
Lett. {\bf 79}, 5238 (1997).

\bibitem{Woldeyohannes99a}
M. Woldeyhannes and S. John, Phys. Rev. A {\bf 60}, 5046 (1999).

\bibitem{Zhu97a}
S.-Y. Zhu, H. Chen and H. Huang, Phys. Rev. Lett. {\bf 79}, 205
(1997).

\bibitem{Zhu00a}
Y.P. Yang and S.-Y. Zhu, Phys. Rev. A {\bf 61}, 043809 (2000).

\bibitem{Zhu00d} Y.P. Yang and S.-Y. Zhu, J. Mod. Opt. {\bf 47}, 1513 (2000).

\bibitem{Paspalakis99a}
E. Paspalakis, N.J. Kylstra and P.L. Knight, Phys. Rev. A {\bf
60}, R33 (1999).

\bibitem{Lee99a}
H. Lee, S.-Y. Zhu and M.O. Scully, Laser Phys. {\bf 9}, 831
(1999).

\bibitem{Erhard00a}
M. Erhard and C.H. Keitel, Opt. Commun. {\bf 179}, 517 (2000).

\bibitem{Angelakis00a}
D.G. Angelakis, E. Paspalakis and P.L. Knight, Phys. Rev. A {\bf
61}, 055802 (2000).

\bibitem{Paspalakis99b}
E. Paspalakis, D.G. Angelakis and P.L. Knight, Opt. Commun. {\bf
172}, 229 (1999).

\bibitem{Zhu00b}
Y.P. Yang, Z.X. Lin, S.-Y. Zhu, H. Chen and W.G. Feng, Phys. Lett.
A {\bf 270}, 41 (2000).

\bibitem{Bay97b}
S. Bay, P. Lambropoulos and K. M{\o}lmer, Phys. Rev. Lett. {\bf
79}, 2654 (1997).

\bibitem{Vats98a}
N. Vats and S. John, Phys. Rev. A {\bf 58}, 4168 (1998).

\bibitem{Nikolopoulos99a}
G.M. Nikolopoulos and P. Lambropoulos, Phys. Rev. A {\bf 60}, 5079
(1999).

\bibitem{Kofman96a}
A.G. Kofman and G. Kurizki, Phys. Rev. A {\bf 54}, R3750 (1996).

\bibitem{Lewenstein99a}
M. Lewenstein and K. Rz\c{a}\.{z}ewski, Phys. Rev. A {\bf 61},
02215 (2000).

\bibitem{Lewenstein88a}
M. Lewenstein, J. Zakrzewski, T.W. Mossberg and J. Mostowski, J.
Phys. B {\bf 21}, L9 (1988); M. Lewenstein, J. Zakrzewski and T.W.
Mossberg, Phys. Rev. A {\bf 38}, 808 (1988).

%\bibitem{Harris92a}
%S.E. Harris, J.E. Field and A. Kasapi, Phys. Rev. A {\bf 46}, R29
%(1992).

%\bibitem{Schmidt96a}
%O. Schmidt, R. Wynands, Z. Hussein and D. Meschede, Phys. Rev. A
%{\bf 53}, R27 (1996).

%\bibitem{Hau99a}
%L.V. Hau, S.E. Harris, Z. Dutton and C.H. Behroozi, Nature {\bf
%397}, 594 (1999).

%\bibitem{Kash99a}
%M.M. Kash, V.A. Sautenkov, A.S. Zibrov, L. Hollberg, G.R. Welch,
%M.D. Lukin, Y. Rostovtsev, E.S. Fry and M.O. Scully, Phys. Rev.
%Lett. {\bf 82}, 5229 (1999).

\bibitem{Paspalakis00a}
E. Paspalakis and P.L. Knight, submitted for publication (2000).

\end{thebibliography}
\end{document}